\documentstyle[12pt,graphicx]{article}

\def\mathfrak{\bf}

\def\be{\begin{equation}}
\def\ee{\end{equation}}
\def\bea{\begin{eqnarray}}
\def\eea{\end{eqnarray}}
\def\dt#1{\on{\hbox{\bf .}}{#1}}                % (big) dot over
\def\Dot#1{\dt{#1}}
\def\IR{\relax{\rm I\kern-.18em R}}
\def\binomial#1#2{\left(\,{\buildrel
{\raise4pt\hbox{$\displaystyle{#1}$}}\over
{\raise-6pt\hbox{$\displaystyle{#2}$}}}\,\right)}

\def\[{\lfloor{\hskip 0.35pt}\!\!\!\lceil}
\def\]{\rfloor{\hskip 0.35pt}\!\!\!\rceil}

\newcommand{\AmS}{{\protect\the\textfont2
  A\kern-.1667em\lower.5ex\hbox{M}\kern-.125emS}}

\catcode`@=11
\def\un#1{\relax\ifmmode\@@underline#1\else
        $\@@underline{\hbox{#1}}$\relax\fi}
\catcode`@=12

\def\fracm#1#2{\hbox{\large{${\frac{{#1}}{{#2}}}$}}}

\def\ad{{\kern0.5pt
                   \alpha \kern-5.05pt
\raise5.8pt\hbox{$\textstyle.$}\kern
0.5pt}}

\def\Dot#1{{\kern0.5pt
     {#1} \kern-5.05pt \raise5.8pt\hbox{$\textstyle.$}\kern
0.5pt}}

\def\a{\alpha}
\def\b{\beta}

\def\d{\delta}
\def\e{\epsilon}

\def\g{\gamma}

\def\k{\kappa}
\def\l{\lambda}

\def\o{\omega}

\def\q{\theta}

\def\s{\sigma}

\def\D{\Delta}

\def\O{\Omega}

\def\S{\Sigma}

\def\ri{{\rm i}}

\def\deb{{\Bar{\de}}}
\newcommand{\non}{\nonumber}
\newcommand{\hf}{\frac12}

\def\cf{{\cal F}}

\def\cm{{\cal M}}

\def\car{{\cal R}}

\def\cy{{\cal Y}}

\def\bo{{\raise.15ex\hbox{\large$\Box$}}}               % D'Alembertian
\def\pa{\partial}                                       % curly d
\def\de{\nabla}                                         % del
\def\TH{{\raise.2ex\hbox{$\displaystyle \bigodot$}\mskip-4.7mu \llap H
\;}}
\def\face{{\raise.2ex\hbox{$\displaystyle \bigodot$}\mskip-2.2mu \llap
{$\ddot
        \smile$}}}                                      % happy face
\def\sp#1{{}^{#1}}                              % superscript(unaligned)
\def\sb#1{{}_{#1}}                              % sub"
\def\Bar#1{\overline{#1}}                       % big bar
\def\leftrightarrowfill{$\mathsurround=0pt \mathord\leftarrow \mkern-6mu
        \cleaders\hbox{$\mkern-2mu \mathord- \mkern-2mu$}\hfill
        \mkern-6mu \mathord\rightarrow$}
\def\dvec#1{\vbox{\ialign{##\crcr
        \leftrightarrowfill\crcr\noalign{\kern-1pt\nointerlineskip}
        $\hfil\displaystyle{#1}\hfil$\crcr}}}           % <--> accent
\def\dt#1{{\buildrel {\hbox{\LARGE .}} \over {#1}}}     % dot-over forsp/sb
\def\fracm#1#2{\hbox{\large{${\frac{{#1}}{{#2}}}$}}}
\def\frac#1#2{{\textstyle{#1\over\vphantom2\smash{\raise.20ex
        \hbox{$\scriptstyle{#2}$}}}}}                   % fraction
\def\sfrac#1#2{{\vphantom1\smash{\lower.5ex\hbox{\small$#1$}}\over
        \vphantom1\smash{\raise.4ex\hbox{\small$#2$}}}} % alternatefraction
\def\bfrac#1#2{{\vphantom1\smash{\lower.5ex\hbox{$#1$}}\over
        \vphantom1\smash{\raise.3ex\hbox{$#2$}}}}       % "
\def\afrac#1#2{{\vphantom1\smash{\lower.5ex\hbox{$#1$}}\over#2}}    % "
\def\on#1#2{\mathop{\null#2}\limits^{#1}}               % arbitraryaccent
\newskip\humongous \humongous=0pt plus 1000pt minus 1000pt
\def\caja{\mathsurround=0pt}
\def\eqalign#1{\,\vcenter{\openup2\jot \caja
        \ialign{\strut \hfil$\displaystyle{##}$&$
        \displaystyle{{}##}$\hfil\crcr#1\crcr}}\,}
\newif\ifdtup

\newcommand{\ve}{\varepsilon}                            %new

  \def\pp{{\mathchoice
          {%displaystyle
              \kern 1pt%
              \raise 1pt
              \vbox{\hrule width5pt height0.4pt depth0pt
                    \kern -2pt
                    \hbox{\kern 2.3pt
                          \vrule width0.4pt height6pt depth0pt
                          }
                    \kern -2pt
                    \hrule width5pt height0.4pt depth0pt}%
                    \kern 1pt
           }
            {%textstyle
              \kern 1pt%
              \raise 1pt
              \vbox{\hrule width4.3pt height0.4pt depth0pt
                    \kern -1.8pt
                    \hbox{\kern 1.95pt
                          \vrule width0.4pt height5.4pt depth0pt
                          }
                    \kern -1.8pt
                    \hrule width4.3pt height0.4pt depth0pt}%
                    \kern 1pt
            }
            {%scriptstyle
              \kern 0.5pt%
              \raise 1pt
              \vbox{\hrule width4.0pt height0.3pt depth0pt
                    \kern -1.9pt  %[e]=0.15pt
                    \hbox{\kern 1.85pt
                          \vrule width0.3pt height5.7pt depth0pt
                          }
                    \kern -1.9pt
                    \hrule width4.0pt height0.3pt depth0pt}%
                    \kern 0.5pt
            }
            {%scriptscriptstyle
              \kern 0.5pt%
              \raise 1pt
              \vbox{\hrule width3.6pt height0.3pt depth0pt
                    \kern -1.5pt
                    \hbox{\kern 1.65pt
                          \vrule width0.3pt height4.5pt depth0pt
                          }
                    \kern -1.5pt
                    \hrule width3.6pt height0.3pt depth0pt}%
                    \kern 0.5pt%}
            }
        }}

  \def\mm{{\mathchoice
                       {%displaystyle
                             \kern 1pt
               \raise 1pt    \vbox{\hrule width5pt height0.4pt depth0pt
                                  \kern 2pt
                                  \hrule width5pt height0.4pt depth0pt}
                             \kern 1pt}
                       {%textstyle
                            \kern 1pt
               \raise 1pt \vbox{\hrule width4.3pt height0.4pt depth0pt
                                  \kern 1.8pt
                                  \hrule width4.3pt height0.4pt depth0pt}
                             \kern 1pt}
                       {%scriptstyle
                            \kern 0.5pt
               \raise 1pt
                            \vbox{\hrule width4.0pt height0.3pt depth0pt
                                  \kern 1.9pt
                                  \hrule width4.0pt height0.3pt depth0pt}
                            \kern 1pt}
                       {%scriptscriptstyle
                           \kern 0.5pt
             \raise 1pt  \vbox{\hrule width3.6pt height0.3pt depth0pt
                                  \kern 1.5pt
                                  \hrule width3.6pt height0.3pt depth0pt}
                           \kern 0.5pt}
                       }}

\def\pd{{\kern0.5pt
                   + \kern-5.05pt \raise5.8pt\hbox{$\textstyle.$}\kern
0.5pt}}

\def\pmd{{\kern0.5pt
                  \pm \kern-5.05pt \raise6.3pt\hbox{$\textstyle.$}\kern1.5pt}}

\def\md{{\mathchoice
   {%displaystyle
      {{\kern 1pt - \kern-6.2pt \raise5pt\hbox{$\textstyle.$}\kern 1pt}}}
    {%textstyle
      {{\kern 1pt - \kern-6.2pt \raise5pt\hbox{$\textstyle.$}\kern 1pt}}}
    {%scriptstyle
      {\kern0.5pt - \kern-5.05pt \raise3.4pt\hbox{$\textstyle.$}\kern0.5pt}}
    {%scriptscriptstyle
      {\kern0.5pt - \kern-5.05pt \raise3.4pt\hbox{$\textstyle.$}\kern0.5pt}}}}

\def\ad{{\dot{\alpha}}}

\def\pp{{\mathchoice
          {%displaystyle
              \kern 1pt%
              \raise 1pt
              \vbox{\hrule width5pt height0.4pt depth0pt
                    \kern -2pt
                    \hbox{\kern 2.3pt
                          \vrule width0.4pt height6pt depth0pt
                          }
                    \kern -2pt
                    \hrule width5pt height0.4pt depth0pt}%
                    \kern 1pt
           }
            {%textstyle
              \kern 1pt%
              \raise 1pt
              \vbox{\hrule width4.3pt height0.4pt depth0pt
                    \kern -1.8pt
                    \hbox{\kern 1.95pt
                          \vrule width0.4pt height5.4pt depth0pt
                          }
                    \kern -1.8pt
                    \hrule width4.3pt height0.4pt depth0pt}%
                    \kern 1pt
            }
            {%scriptstyle
              \kern 0.5pt%
              \raise 1pt
              \vbox{\hrule width4.0pt height0.3pt depth0pt
                    \kern -1.9pt  %[e]=0.15pt
                    \hbox{\kern 1.85pt
                          \vrule width0.3pt height5.7pt depth0pt
                          }
                    \kern -1.9pt
                    \hrule width4.0pt height0.3pt depth0pt}%
                    \kern 0.5pt
            }
            {%scriptscriptstyle
              \kern 0.5pt%
              \raise 1pt
              \vbox{\hrule width3.6pt height0.3pt depth0pt
                    \kern -1.5pt
                    \hbox{\kern 1.65pt
                          \vrule width0.3pt height4.5pt depth0pt
                          }
                    \kern -1.5pt
                    \hrule width3.6pt height0.3pt depth0pt}%
                    \kern 0.5pt%}
            }
        }}

  \def\mm{{\mathchoice
                       {%displaystyle
                             \kern 1pt
               \raise 1pt    \vbox{\hrule width5pt height0.4pt depth0pt
                                  \kern 2pt
                                  \hrule width5pt height0.4pt depth0pt}
                             \kern 1pt}
                       {%textstyle
                            \kern 1pt
               \raise 1pt \vbox{\hrule width4.3pt height0.4pt depth0pt
                                  \kern 1.8pt
                                  \hrule width4.3pt height0.4pt depth0pt}
                             \kern 1pt}
                       {%scriptstyle
                            \kern 0.5pt
               \raise 1pt
                            \vbox{\hrule width4.0pt height0.3pt depth0pt
                                  \kern 1.9pt
                                  \hrule width4.0pt height0.3pt depth0pt}
                            \kern 1pt}
                       {%scriptscriptstyle
                           \kern 0.5pt
             \raise 1pt  \vbox{\hrule width3.6pt height0.3pt depth0pt
                                  \kern 1.5pt
                                  \hrule width3.6pt height0.3pt depth0pt}
                           \kern 0.5pt}
                       }}

\def\pd{{\kern0.5pt
                   + \kern-5.05pt \raise5.8pt\hbox{$\textstyle.$}\kern
0.5pt}}

\def\pmd{{\kern0.5pt
                  \pm \kern-5.05pt \raise6.3pt\hbox{$\textstyle.$}\kern1.5pt}}

\def\md{{\mathchoice
   {%displaystyle
      {{\kern 1pt - \kern-6.2pt \raise5pt\hbox{$\textstyle.$}\kern 1pt}}}
    {%textstyle
      {{\kern 1pt - \kern-6.2pt \raise5pt\hbox{$\textstyle.$}\kern 1pt}}}
    {%scriptstyle
      {\kern0.5pt - \kern-5.05pt \raise3.4pt\hbox{$\textstyle.$}\kern0.5pt}}
    {%scriptscriptstyle
      {\kern0.5pt - \kern-5.05pt \raise3.4pt\hbox{$\textstyle.$}\kern0.5pt}}}}

\def\dslash{\not{\hbox{\kern-2pt $\partial$}}}
\def\Dslash{\not{\hbox{\kern-4pt $D$}}}
\def\pslash{\not{\hbox{\kern-2.3pt $p$}}}
 \newtoks\slashfraction
 \slashfraction={.13}
 \def\slash#1{\setbox0\hbox{$ #1 $}
 \setbox0\hbox to \the\slashfraction\wd0{\hss \box0}/\box0 }

\font\ro=cmsy10                          % font with rope
\def\kcr{{\hbox{\ro \char'170}}}                % right-handed rope
\def\ktl{{\hbox{\ro \char'170}}}        % top end for left-handed rope
\def\ktr{{\hbox{\ro \char'170}}}        % " right
\def\kbl{{\hbox{\ro \char'170}}}        % " bottom left
\def\kbr{{\hbox{\ro \char'170}}}        % " right
\def\plpl{\raise-2pt\hbox{$\raise3pt\hbox{$_+$}\hskip-6.67pt\raise0.0pt
\hbox{$^+$}\hskip 0.01pt$}}
\def\mimi{\raise-2pt\hbox{$\raise3pt\hbox{$_-$}\hskip-6.67pt\raise0.0pt
\hbox{$^-$}\hskip 0.01pt$}}

\def\bo{{\raise.15ex\hbox{\large$\Box$}}}               % D'Alembertian
\def\pa{\partial}                                       % curly d
\def\de{\nabla}                                         % del
\def\TH{{\raise.2ex\hbox{$\displaystyle \bigodot$}\mskip-4.7mu \llap H \;}}
\def\face{{\raise.2ex\hbox{$\displaystyle \bigodot$}\mskip-2.2mu \llap {$\ddot
        \smile$}}}                                      % happy face
\def\sp#1{{}^{#1}}                              % superscript (unaligned)
\def\sb#1{{}_{#1}}                              % sub"
\def\Bar#1{\overline{#1}}                       % big bar
\def\leftrightarrowfill{$\mathsurround=0pt \mathord\leftarrow \mkern-6mu
        \cleaders\hbox{$\mkern-2mu \mathord- \mkern-2mu$}\hfill
        \mkern-6mu \mathord\rightarrow$}
\def\dvec#1{\vbox{\ialign{##\crcr
        \leftrightarrowfill\crcr\noalign{\kern-1pt\nointerlineskip}
        $\hfil\displaystyle{#1}\hfil$\crcr}}}           % <--> accent
\def\dt#1{{\buildrel {\hbox{\LARGE .}} \over {#1}}}     % dot-over for sp/sb
\def\fracm#1#2{\hbox{\large{${\frac{{#1}}{{#2}}}$}}}
\def\frac#1#2{{\textstyle{#1\over\vphantom2\smash{\raise.20ex
        \hbox{$\scriptstyle{#2}$}}}}}                   % fraction
\def\sfrac#1#2{{\vphantom1\smash{\lower.5ex\hbox{\small$#1$}}\over
        \vphantom1\smash{\raise.4ex\hbox{\small$#2$}}}} % alternate fraction
\def\bfrac#1#2{{\vphantom1\smash{\lower.5ex\hbox{$#1$}}\over
        \vphantom1\smash{\raise.3ex\hbox{$#2$}}}}       % "
\def\afrac#1#2{{\vphantom1\smash{\lower.5ex\hbox{$#1$}}\over#2}}    % "
\def\on#1#2{\mathop{\null#2}\limits^{#1}}               % arbitrary accent
\parskip=\medskipamount                 % space between paragraphs (LaTeX)
\thicklines                         % thick straight lines for pictures (LaTeX)

\def\oldheadpic{                                % old UM heading
        \setlength{\unitlength}{.4mm}
        \thinlines
        \par
        \begin{picture}(349,16)
        \put(325,16){\line(1,0){4}}
        \put(330,16){\line(1,0){4}}
        \put(340,16){\line(1,0){4}}
        \put(335,0){\line(1,0){4}}
        \put(340,0){\line(1,0){4}}
        \put(345,0){\line(1,0){4}}
        \put(329,0){\line(0,1){16}}
        \put(330,0){\line(0,1){16}}
        \put(339,0){\line(0,1){16}}
        \put(340,0){\line(0,1){16}}
        \put(344,0){\line(0,1){16}}
        \put(345,0){\line(0,1){16}}
        \put(329,16){\oval(8,32)[bl]}
        \put(330,16){\oval(8,32)[br]}
        \put(339,0){\oval(8,32)[tl]}
        \put(345,0){\oval(8,32)[tr]}
        \end{picture}
        \par
        \thicklines
        \vskip.2in}
\def\oldtitle#1#2#3#4{\oldheadpic\begin{center}\vglue.5in{\large\bf #1}\\[.6in]
        {#2}\\[.1in] {\it Department of Physics and Astronomy}\\
        {\it University of Maryland, College Park, MD 20742}\\[.6in]
        Physics Publication \#{#3}\\ {#4}\\[1.5in] {\bf ABSTRACT}\\[.1in]
        \end{center} \begin{quotation}}                 % old title stuff
\def\oldTitle#1#2#3#4#5#6#7{\oldheadpic\begin{center} \vglue .4in
        {\large\bf #1}\\[.4in]
        {#2}\\[.1in] {\it Department of Physics and Astronomy}\\
        {\it University of Maryland, College Park, MD 20742}\\[.1in]
        {#3}\\[.1in] {\it {#4}}\\ {\it {#5}}\\[.4in]
        Physics Publication \#{#6}\\ {#7}\\[.5in] {\bf ABSTRACT}\\[.1in]
        \end{center} \begin{quotation}}                 % " for 2 authors
\def\border{                                            % border
        \setlength{\unitlength}{1mm}
        \newcount\xco
        \newcount\yco
        \xco=-21
        \yco=12
        \begin{picture}(140,0)
        \put(\xco,\yco){$\ktl$}
        \advance\yco by-1
        {\loop
        \put(\xco,\yco){$\kcr$}
        \advance\yco by-2
        \ifnum\yco>-240
        \repeat
        \put(\xco,\yco){$\kbl$}}
        \xco=158
        \yco=12
        \put(\xco,\yco){$\ktr$}
        \advance\yco by-1
        {\loop
        \put(\xco,\yco){$\kcr$}
        \advance\yco by-2
        \ifnum\yco>-240
        \repeat
        \put(\xco,\yco){$\kbr$}}
        \put(-20,13){\tiny **University of Maryland * Center for String and
         Particle  Theory* Physics Department***University of Maryland *Center
        for String and Particle  Theory** }
        \put(-20,-241.5){\tiny **University of Maryland * Center for String and
         Particle  Theory* Physics Department***University of Maryland *Center
        for String and Particle  Theory** }
        \end{picture}
        \par\vskip-8mm}
\def\bordero{                                           % alternate border
        \setlength{\unitlength}{1mm}
        \newcount\xco
        \newcount\yco
        \xco=-31
        \yco=12
        \begin{picture}(140,0)
        \put(\xco,\yco){$\ktl$}
        \advance\yco by-1
        {\loop
        \put(\xco,\yco){$\kclr}
        \advance\yco by-2
        \ifnum\yco>-240
        \repeat
        \put(\xco,\yco){$\kbl$}}
        \xco=151
        \yco=12
        \put(\xco,\yco){$\ktr$}
        \advance\yco by-1
        {\loop
        \put(\xco,\yco){$\kcr$}
        \advance\yco by-2
        \ifnum\yco>-240
        \repeat
        \put(\xco,\yco){$\kbr$}}
        \put(-20,12){\ooo bacdefghidfghghdhededbihdgdfdfhhdheidhdhebaaahjhhdahba

hgdedge
   hgfdiehhgdigicba}
        \put(-20,-241.5){\ooo ababaighefdbfghgeahgdfgafagihdidihiidhiagfedhadbfd

ecdcdfa
   gdcbhaddhbgfchbgfdacfediacbabab}
        \end{picture}
        \par\vskip-8mm}
\def\headpic{                                           % UM heading
        \indent
        \setlength{\unitlength}{.4mm}
        \thinlines
        \par
        \begin{picture}(29,16)
        \put(165,16){\line(1,0){4}}
        \put(170,16){\line(1,0){4}}
        \put(180,16){\line(1,0){4}}
        \put(175,0){\line(1,0){4}}
        \put(180,0){\line(1,0){4}}
        \put(185,0){\line(1,0){4}}
        \put(169,0){\line(0,1){16}}
        \put(170,0){\line(0,1){16}}
        \put(179,0){\line(0,1){16}}
        \put(180,0){\line(0,1){16}}
        \put(184,0){\line(0,1){16}}
        \put(185,0){\line(0,1){16}}
        \put(169,16){\oval(8,32)[bl]}
        \put(170,16){\oval(8,32)[br]}
        \put(179,0){\oval(8,32)[tl]}
        \put(185,0){\oval(8,32)[tr]}
        \end{picture}
        \par\vskip-6.5mm
        \thicklines}
\def\title#1#2#3#4{\border\headpic {\hbox to\hsize{#4 \hfill UMDEPP #3}}\par
        \begin{center} \vglue .5in {\large\bf #1}\\[.6in]
        {#2}\\[.1in] {\it Department of Physics and Astronomy}\\
        {\it University of Maryland, College Park, MD 20742}\\[1.5in]
        {\bf ABSTRACT}\\[.1in] \end{center} \begin{quotation}}  % title stuff
\def\Title#1#2#3#4#5#6#7{\border\headpic
        {\hbox to\hsize{#7 \hfill UMDEPP #6}}\par
        \begin{center} \vglue .4in {\large\bf #1}\\[.4in]
        {#2}\\[.1in] {\it Department of Physics and Astronomy}\\
        {\it University of Maryland, College Park, MD 20742}\\[.1in]
        {#3}\\[.1in] {\it {#4}}\\ {\it {#5}}\\[.5in] {\bf ABSTRACT}\\[.1in]
        \end{center} \begin{quotation}}                 % " for 2 authors
\def\endtitle{\end{quotation}\newpage}                  % end title page

\def\qd{{\kern0.5pt
                   q \kern-5.05pt \raise5.8pt\hbox{$\textstyle.$}\kern
0.5pt}}

\begin{document}

\def\dt#1{\on{\hbox{\bf .}}{#1}}                % (big) dot over
\def\Dot#1{\dt{#1}}

\def\gfrac#1#2{\frac {\scriptstyle{#1}}
        {\mbox{\raisebox{-.6ex}{$\scriptstyle{#2}$}}}}
\def\gg{{\hbox{\sc g}}}
\border\headpic {\hbox to\hsize{July 2009 \hfill
{UMDEPP 09-043}}}
\par
{$~$ \hfill
{hep-th/0907.5264}}
\par

\setlength{\oddsidemargin}{0.3in}
\setlength{\evensidemargin}{-0.3in}
\begin{center}
\vglue .10in
{\large\bf Ectoplasm \& Superspace Integration Measure for \\
\vskip.1in
$2$D Supergravity with Four Spinorial Supercurrents\footnote
{Supported in part  by National Science Foundation Grant
PHY-0354401.}\  }
\\[.5in]

S.\, James Gates, Jr.\footnote{gatess@wam.umd.edu}
and Gabriele Tartaglino-Mazzucchelli\footnote{gtm@umd.edu} 
\\[0.2in]

{\it Center for String and Particle Theory\\
Department of Physics, University of Maryland\\
College Park, MD 20742-4111 USA}\\[1.5in]

{\bf ABSTRACT}\\[.01in]
\end{center}
\begin{quotation}
{Building on a previous derivation of the local chiral projector for a two dimensional
superspace with eight real supercharges, we provide the complete density projection
formula required for locally supersymmetrical theories in this context.  The derivation
of this result is shown to be very efficient using techniques based on the Ectoplasmic
construction of local measures in superspace.}

${~~~}$ \newline

\endtitle

\section{Introduction}

~~~~ Some years ago, a formulation of a 2D supergravity theory which included
off-shell closure of the local supersymmetry algebra with four real spinorial supercharges 
and a necessary set of auxiliary fields was introduced into the literature \cite{2DN4SG}.  
In a subsequent development there was made a proposal (called `Ectoplasm' \cite{Ecto}) 
for a conceptual framework leading to efficient derivations of local superspace 
integration measures (density projection operators)\footnote{
A mathematical construction giving the formal bases for the Ectoplasm methods 
can be found \\ $~~~$$~~~$ in the theory of integration over surfaces in 
supermanifolds developed in \cite{GKSchwarz,BS,Vor}.}.  In addition, about the same time 
there was put forward an alternative general framework for the derivation of density projection 
operators based on the use of superspace normal coordinate expansions first introduced 
in \cite{McA} and rediscovered\footnote{Previous approaches for component reduction, ultimately 
related to normal coordinates \\ $~~~$$~~~$ expansions, can be found in 
\cite{WZ,LR,Muller82,Ramirez,Muller,Muller89,AD}.}
 in \cite{Norcor};
see \cite{KuzGtmNormCoord} for recent reformulations and improvements of the normal 
coordinates techniques.
%\footnote{For some recent developments 
%and applications of normal coordinates techniques see also \cite{KuzGtmNormCoord}.}.
The ectoplasm and normal coordinates frameworks have been found to be closely related 
\cite{Ectonorcor,Fields}.

Prior to the introduction of the ectoplasmic and normal coordinate approaches, the
question of how to construct local superspace supergravity densities had been approached
by two other and more cumbersome methods.  Both of these can be seen in two books 
on the subject. In the first, {\em {Superspace}} \cite{ggrs1}, an approach was taken 
to reproduce, at the level of superfields, a Noether approach thus leading to the density
projector.  In the second {\em {Ideas}} \cite{Ideas}, an approach that was taken to 
utilize the prepotential formulation of supergravity theory to derive the density
projector.

It has been argued from its inception that the ectoplasmic concept is not only extremely
efficient but also likely applies to even more complicated theories such as string theory. 
Though there was no such evidence at the time of the introduction of the ectoplasm
approach, later it was shown that integration measures in the `pure spinor 
formulation' of superstrings follow precisely from the extension of the ectoplasmic concept
 to this realm of theories \cite{Pspinors}.

The off-shell formulation of a 2D, $\cal N$ = 4  supergravity theory implies  the
existence of a straightforward way to completely develop an {\em {efficient}} local integration 
theory for the associated local Salam-Strathdee superspace.  We will complete such a 
construction in the current work by use of the ectoplasmic suggestion.

%%%
This paper is organized as follows.
In section 2 we review the 2D, $\cal N$=4 supergravity formulation of \cite{2DN4SG}.
Section 3 is devoted to the presentation of a new super 2-form multiplet.
In section 4 we make use of the ectoplasmic approach to build the density projector for
 the 2D, $\cal N$=4 supergravity of \cite{2DN4SG}; this is the main result of the paper.
Section 5 collects some conclusions.
The paper includes two appendices. Appendix A contains the derivation of the result of section 3.
Then, the appendix B is a collection of formulas used in the paper.

%%%%%%%%%%%%%%%%%%%%%%%%%%%%%%%%%%%%%%%%%%%
\section{An Off-Shell 2D Supergravity Geometry With Eight Real Local Supersymmetries}

~~~~In this section we review some aspects of the off-shell 2D, ${\cal N} =4$
 minimal supergravity
multiplet first introduced in \cite{2DN4SG}. We focus on the curved superspace geometry
underlining the minimal supergravity that will be used in the computations of this paper. 

%~~~~ 
The work in \cite{2DN4SG} showed there exists 
component fields $ (e_a{}^m ,~ \psi_a {}^{\a i}, ~ A_{ai} {}^j , ~ B,  ~ G, ~H ) $ which describe
an off-shell 2D supergravity theory possessing eight real local (or four real spinorial) 
supercharges.  The previous list of component fields contains the graviton, the gravitini, SU(2) connection, a complex scalar $B$, one real scalar $G$ and one real pseudoscalar $H$.
 These are the components associated
with the following constraints on the 2D, ${\cal N}$ = 4 superspace supergravity covariant 
derivative algebra\footnote{In the present paper we adopt the Lorentz and SU(2) notations collected in Appendix   \\ $~~~$$~~~$
A of \cite{GM} and consistent with the conventions of \cite{ggrs1}.}
\bea
\{\, \de \sb{\a i} ~ ,~  \de \sb{\b j} \,  \} & = & 2 \Bar B [ \, C \sb {\a \b}
C \sb{ ij} \cm ~-~ ( \g \sp 3 ) \sb{ \a \b} \cy \sb{i j} \, ] \,  ~~, 
{~~~~~~~~~} {~~~~~~~~~} {~~~~~~~~~} {~~~~~~~~~~~}
   \label{CommAlg1}
\\
\{  \,  \deb_{\a}{}^i ~,~ \deb_{\b}{}^j \, \, \} & = & 2 B {[} \, C_{\a \b}
C^{ ij} \cm ~-~ ( \g \sp 3 )_{ \a \b} \cy^{i j} \, {]} \,  ~~,
{~~~~~~~~~} {~~~~~~~~~} {~~~~~~~~~} {~~~~~~~~~~~}
    \label{CommAlg2}
\\
\{ \, \de \sb{\a i} ~,~ \Bar \de \sb{\b} \sp j  \, \}   &=&
2\ri\,  \d \sb i \sp j (\g \sp{c}) \sb{\a \b} \de \sb{c} 
~+~ 2 \d \sb i \sp j \phi \sb{\a}{}^{\g}(\g^3)_{\g \b}  \cm  
~-~ 2\phi \sb{\a \b} \cy \sb i \sp j  ~~,    {~~~~~~~~}     {~~~~~~~~}
    \label{CommAlg3}
\\
{[} \, \de \sb{\a i} ~,~ \de \sb{b} \, {]}   &=& 
\fracm \ri2 \phi \sb {\a} \sp{\g}(\g \sb b) \sb \g \sp {\b} \de \sb{\b i} 
~+~ \fracm \ri2 (\g^3 \g \sb b)\sb {\a} \sp{ \b} \Bar B C \sb{i j} \Bar \de \sb \b \sp j
   \non\\     
&&
~-~ \ri(\g^3 \g_b)_{\a \b}\bar{\S}^{\b}{}_i \cm 
~+~ \ri (\g \sb b)_{\a \b}\bar{\S}^{\b}{}_j\cy_{i}{}^{j} ~~,  
  {~~~~~~~~~~}     {~~~~~~~~~~}   {~~~~~~~~}    
   \label{CommAlg4}
\\
~{[} \, \deb_{\a}{}^{i} ~,~ \de \sb{b} \,{]}   &=& 
-~\fracm \ri2 \bar{\phi}_{\a}{}^{\g}(\g_b)_\g{}^{\b} \deb_{\b}{}^{i} 
~+~  \fracm \ri2 (\g^3 \g_b)_{\a}{}^{\b} B C^{i j} \de_{\b j}
   \non\\     
&~&~~ 
-~ \ri(\g^3 \g_b)_{\a \b}\S^{\b i} \cm 
~-~ \ri (\g \sb b)_{\a \b}\S^{\b j}\cy^{i}{}_{j} ~~,  
  {~~~~~~~~}     {~~~~~~~~~~}   {~~~~~~~~}   
   \label{CommAlg5}
\\
~{[}  \, \de \sb{a} ~,~ \de \sb{b} \,{]}  &=& - ~\fracm 12 \ve_{a b} 
[ (\g \sp 3) \sb \a \sp \b \S \sp{ \a i} \de \sb{ \b i}
~+~ (\g \sp 3) \sb \a \sp \b \Bar \S \sp \a \sb i \Bar \de \sb \b \sp i
~+~ \car \cm ~+~ \ri \cf \sb i \sp j \cy \sb j \sp i ]~ .   ~~~~
 \label{CommAlg6}
\eea
where
\bea 
(B)^*&=&\bar{B}~,~~~
(G)^*~=\, G~,~~~
(H)^*~=\, H~,~~~
(\S_\a{}^i)^*~=\,\bar{\S}_{\a i}~,
\\
\phi_{\a \, \b}& =&C_{\a \b} G +\ri (\g^3)_{\a \b} H~,
\\
\bar{\phi}_{\a \, \b}&=&(\phi_{\a\b})^*=- C_{\a \b} G +\ri (\g^3)_{\a \b} H =\phi_{\b\a}~.
\eea
In writing these, we have corrected some coefficients that appear in the work
of \cite{GM} in the terms that appear in (\ref{CommAlg3}) - (\ref{CommAlg5}).
These corrected coefficients do not affect (\ref{CommAlg1}) and  (\ref{CommAlg2}).
Thus the result in the work of \cite{GM} is unaffected by this change.

In the previous algebra, the covariant derivatives are $\de_A=(\de_a,\de_{\a i},\deb_\a{}^i)$
\bea
\de_A~=~
E_A{}^M\pa_M+\o_A\cm+\ri\,\Gamma_A{}_k{}^l\cy_l{}^k~.
\label{covDev}
\eea
The 2D, ${\cal N}=4$  curved superspace is locally parametrized by the coordinates
$z^M=(x^m,\q^{\mu i},\bar{\q}^\mu{}_i)$  with the Grassmann variables 
$\q^{\mu i}$ and 
$\bar{\q}^\mu{}_i$ related by complex conjugation $\bar{\q}^\mu{}_i=(\q^{\mu i})^*$; the bosonic 
coordinates will be also denoted as $x^m=(\tau,\sigma)$.
In (\ref{covDev}), $E_{A}{}^M$ is the inverse of the vielbein 
$E_M{}^A$ ($E_M{}^AE_A{}^N=\d_M^N$, $E_A{}^ME_M{}^B=\d_A^B$)
with $\pa_M=\pa/\pa z^M$,
$\o_A$ the 2D Lorentz connection and $\Gamma_A{}_{k}{}^l$ is the SU(2) connection.
The torsion $T_{AB}{}^C$, Lorentz curvature $R_{AB}$ and SU(2) curvature 
$R_{AB}{}_k{}^l$ superfields are defined by (1) - (6) and
\bea
[\de_A,\de_B\}~=~
T_{AB}{}^C\de_C+R_{AB}\cm+\ri\,R_{AB}{}_k{}^l\cy_l{}^k~.
\eea
The action of the local 2D Lorentz generator $\cm$ and 
of the local SU(2) generator 
$\cy_{k}{}^{l}$ on the spinor covariant derivatives are the following ($\cy_{kl}=\cy_k{}^pC_{pl}$)
\bea
{[}\cm,\de_{\a i}{]}&=&\hf(\g^3)_\a{}^\b\de_{\b i}~,~~~
{[}\cm,\deb_{\a}{}^{i}{]}=\hf(\g^3)_\a{}^\b\deb_{\b}{}^{i}~,~~~
\\
{[}\cy_{kl},\de_{\a i}{]}&=&\hf C_{i(k}\de_{\b l)}~,~~~
{[}\cy_{kl},\deb_{\a}{}^{i}{]}=-\hf \d^i_{(k}\deb_{\b l)}~.
\eea

It is worthy  to recall that the consistency of the Bianchi identities constructed from the commutator
algebra above requires the conditions \cite{2DN4SG}
\bea
\Bar \de \sb \a \sp i B &=&0~~~~~~,~~~
\de \sb{\a i} B~ =\, -2 C \sb{i j} ( \g \sp 3 ) \sb{ \a \b} \S \sp{ \b j}~~,
 \\
\de \sb{\a i} G &=& \Bar \S \sb{ \a i}~~~,~~~  
\de \sb{\a i} H~ =\, \ri( \g \sp 3 ) \sb{ \a} \sp{ \b} \Bar \S \sb{ \b i},
~~~, \\
\Bar \de \sb \a \sp i \S \sp{\b j} &=& \ri  C \sp{i j}
(\g \sp 3 \g \sp a) \sb {\a } \sp{ \b} \de \sb a B  ~~,  \\
\de \sb{\a i} \S \sp{\b j} &=& {1 \over 2} \d \sb \a \sp \b
\d \sb i \sp j [ \car ~-~ 2 G \sp 2 ~-~ 2 H \sp 2 ~-~ 2 B \Bar B ]
~+~ \ri (\g \sp 3) \sb \a \sp \b \cf \sb i \sp j  \\
& &+~ \ri \d \sb i \sp j (\g \sp a ) \sb \a \sp \b (\de \sb a G) 
-~  \d \sb i \sp j (\g \sp 3 \g \sp a ) \sb \a \sp \b 
(\de \sb a H) ~~~.    
 \label{BIs}
\eea

The component gauge fields occur in the above supertensors in the 
following manner\footnote{Given a superfield $\Psi(\tau,\sigma,\q,\bar{\q})$, we denote as 
usual with 
$\Psi|:=\Psi|_{\q=0}$ the field obtained by \\ $~~~$$~~~$
 setting to zero all the Grassmanian coordinates.}
\begin{equation}
\begin{array}{lll} 
\car {\big |} &=& \ve^{a b} {\large \{ } ~ {\car}_{a b} (\hat \omega) ~+~  
[ ~ 2\ri (\g^3 \g_a) _{\a \b} \psi_b {}^{\a i} \Bar \S^\b {}_i ~+~ {\rm 
{h.c.}} ~  ]  \\ 
& &~~~~~~ +~ 4 \phi_\a {}^\g (\g^3)_{\g \b} \psi_a {}^{\a i} {\Bar \psi}_b 
{}^\b {}_i  ~-~2  [~ C_{i j} \Bar B \psi_a {}^{\a i}  \psi_{b \a}{}^{j}
~+~ {\rm {h.c.}} ~  ]~ {\large \} } ~~, \\
&   & \\
\S ^{\a i} {\big |} & = & \ve^{ab} {\large \{ } ~ \psi_{ab} {}^{\b i}
(\g^3)_\b {}^\a ~-~ \ri \psi_a {}^{\b i} {\phi}_\b {}^\g 
( \g^3 \g_b)_\g {}^\a   ~+~ \ri C^{ij} B \Bar \psi_a {}^\b {}_j 
(\g_b)_\b {}^\a  ~ {\large \} } ~~, \\ 
&   & \\
\cf_i {}^j {\big |} &=& \ve^{ab} {\large \{ } ~ {\rm F}_{a b}(A)_{i}{}^j 
~-~  2\ri  (\g_a)_{\a \b}  [~ \psi_b {}^{\a j} \Bar \S^\b {}_i ~+~ \Bar \psi_b 
{}^\a {}_i \S^{\b j} \\ 
& &~~~~~~~~~~~~~~~~~~-~ \frac 12 \d_i^j (\psi_b {}^{\a k} \Bar \S^\b {}_k 
~+~  \Bar \psi _b {}^\a {}_k \S^{\b k}) ~] \\
& & ~~~~~~~~~~~~~~~~~~-~ 4  \phi_{\a \b} [\psi_a {}^{\a j} \Bar
\psi_b{}^\b{}_i ~-~ \frac 12 \d_i^j \psi_a {}^{\a k} \Bar
\psi_b{}^\b{}_k ] \\             
& & ~~~~~~~~~~~~~~~~~~-~ 2 (\g^3)_{\a \b} [~\Bar B ( C_{i k} \psi_a {}^{\a k} 
\psi_b {}^{\b}{}^{k} ~-~ \frac 12 \d_i^j C_{k l} \psi_a {}^{\a k} \psi_b 
{}^{\b l})
\\ & & ~~~~~~~~~~~~~~~~~~+~ B ( C^{j k} \Bar \psi_a {}^{\a}{}_{i}
 \Bar \psi_{b}{}^{ \b}{}_{k} ~-~ \frac 12 \d_i^j  C^{k l} \Bar \psi_a{}^{\a}
{}_{k} \Bar \psi_{b}{}^{\b}{}_{l})~ ] ~{\large \} } ~~ ,   \end{array}      
\label{FSs}  
\end{equation}
where $\ve^{a b} \, {\car}_{a b} (\hat \omega)$ is the usual two-dimensional curvature in
terms of the inverse of the vielbein ${e}_a {}^m$ and of the Lorentz connection $\hat \omega_a $;
$ \ve^{ab}\psi_{ab} {}^{\b i}$ is the gravitini field strength;
$\ve^{ab}F_{ab}(A)$ is the SU(2) field strength function of ${e}_a {}^m$ and of the SU(2)
connection $A_a{}_k{}^l$ \cite{2DN4SG}.
The component gauge fields $e_a{}^m,\,\hat{\o}_a,\, A_{a}{}_k{}^l$ are easily related to the 
gauge superfields $E_A{}^M,\,\o_A,\,\Gamma_A{}_k{}^l$ in (\ref{covDev}) by using standard
Wess-Zumino gauge reduction techniques \cite{ggrs1,Ideas}.

%%%%%%%%%%%%%%%%%%%%%%%%%%%%%%%%%%%%%%%%%%
\section{Defining A Closed 2D, ${\cal N}=4$  Super Two-Form }

~~~~
In this section we are going to present a new closed 2D, ${\cal N}=4$ super two-form defined in 
terms of an unconstrained scalar chiral superfield. The result contained in the Theorem 1 is crucial 
to build the measure of the local superspace integration theory for 2D, ${\cal N}=4$ supergravity
theories as we will see in section 4.

The work in \cite{GM} 
established that the fourth-order spinorial derivatives operator $ {\cal {\Bar D}}{}^{(4)} $, defined by
\be \eqalign{
 {\cal {\Bar D}}{}^{(4)} ~=~ 
 \left[ \, {\Bar {\nabla}}{}^{(2) \, \a \, \b} 
 ~-~ 2 \, B \, (\g^3){}^{\a \, \b} \, \right] \, {\Bar {\nabla}}{}^{(2)}_{ \, \a \, \b} 
~~, } \label{ChRLprj}  \ee
is the chiral projection operator satisfying 
\be \eqalign{ {\Bar \nabla}{}_{\g}^{ i} \,   {\cal {\Bar D}}{}^{(4)}   \, \Psi 
~=~ {\Bar \nabla}{}_{\g}^{ i} \,  \left[ \, {\Bar {\nabla}}{}^{(2) \, \a \, \b}  ~-~ 2 \, B \, 
(\g^3){}^{\a \, \b} \, \right] \, {\Bar {\nabla}}{}^{(2)}_{ \, \a \, \b} \, \Psi  ~=~ 0
 } \label{ChRLcond}   \ee
for any general scalar superfield $ \Psi$.  We note the derivation of 
(\ref{ChRLprj}) and (\ref{ChRLcond}) given in \cite{GM}
follows solely from algebraic manipulations of the derivatives that appear in (\ref{CommAlg2}).

In a later section we will exploit the fact that a closed 2D, $\cal N$ = 4 super two-form is 
sufficient to determine the local integration measure for an appropriate curved superspace.
For this purpose it is necessary to define the components of a 2D, $\cal N$ = 4 
super two-form.  The general framework for the construction of such forms was presented
some time ago \cite{pForm} which implies for the present consideration we should introduce
a super 2-form whose component superfields may be written in the form $  J_{A}
{}_{B} ~=~ \big(\,  J_{\a i}{}_{\b j} , ~  J_{\a  i}{}_{\b}{}^{j} ,~  J_{\a}{}^i{}_{\b}{}^j , ~ 
J_{\g k}{}_{a}, ~ J_{\g}{}^k{}_{a} , ~ J_{ab}\, \big)$.
We refer the reader to \cite{pForm,ggrs1} for the notations we adopt in the use of super
p-forms.
In general, given a super p-form $\O$, described by the component superfields 
$\O_{A_1\cdots A_p}$, its exterior derivative $F=d\O$ has components  $F_{A_1\cdots A_{p+1}}$
given by
\footnote{With $[\cdots)$ we denote the complete graded symmetrization of indeces.}
\bea
F_{A_1\cdots A_pA_{p+1}}={1\over p!}\de_{[A_1}\O_{A_2\cdots A_{p+1})}
-{1\over 2((p-1)!)}T_{[A_1A_2|}{}^B\O_{B|A_3\cdots A_{p+1})}~.
\eea
The superform $\O$ is closed if 
$F_{A_1\cdots A_{p+1}}=0$.
We can now state a theorem. 
%\vskip.05in 
%\noindent
~\\~\\~\\
$~~~${\it {Theorem 1}} \newline  \noindent
$~~~$ {\it {If $U$} is a chiral superfield, i.e. satisfies} $\,{\Bar \nabla}{}_{\a}^{ i} \, U 
\, = \, 0\, ,$ {\it {the components defined by}}
 \be \eqalign{
 J_{\a i}{}_{\b}{}^{ j} &=~ 0 
~~~,  \cr
%%%%%%%%%%%%%%%%%%%%%%%%%%%%%%%%%%
 J_{\a i}{}_{\b j} &=~
2(\g^3)_{\a\b}\deb^{(2)}_{ij}\Bar{U}
-\,C_{\a\b}C_{ij}(\g^3)^{\g\d}\deb^{(2)}_{\g\d}\Bar{U}
~~~,  \cr
%%%%%%%%%%%%%%%%%%%%%%%%%%%%%%%%%%
 J_{\a}{}^i{}_{\b}{}^j &=~
 \, 2 \, (\g^3)_{\a\b}\de^{(2) \,ij}{U}
~-~ \,C_{\a\b}C^{ij}(\g^3)^{\g\d}\de^{(2)}_{\g\d}{U}
~~~,  \cr
%%%%%%%%%%%%%%%%%%%%%%%%%%%%%%%%%%
J_{\g k} {}_{a}&=~ 
-~\fracm \ri3\ve_{ab}(\g^b)_{\g}{}^{\d}\deb_\d {}^{p}\deb^{(2)}_{kp}\Bar{U}
~~~, \cr
%%%%%%%%%%%%%%%%%%%%%%%%%%%%%%%%%%
J_{\g}{}^k{}_{a}
&=~ -~
\fracm \ri3 \ve_{ab}(\g^b)_{\g}{}^{\d}\de_{\d p}\de^{(2) \,kp}{U}
~~~,  \cr
%%%%%%%%%%%%%%%%%%%%%%%%%%%%%%%%%%
J_{ab}
&=~
-~\fracm 18 \,  \ve_{ab}\, \Big{[} \,\Big(
\de^{(4)} ~-~ 2 \, \bar{B}(\g^3)^{\a\b}\de^{(2)}_{\a\b}\Big){U}
~+~
\Big( \deb^{(4)} ~-~ 2 \,
B(\g^3)^{\a\b}\deb^{(2)}_{\a\b}\Big)\Bar{U} 
\,\Big{]}
~,
  } \label{2Form}   \ee
 \newline $~~~$
{\it {describe a closed 2D, $\cal N$ = 4 super two-form with respect to the supergravity}}
 \newline $~~~$
{\it {commutator algebra in Eq. (\ref {CommAlg1}) - Eq. (\ref {CommAlg6}).}} 
\vskip.2pt
\noindent
The superfield $\bar{U}:=(U)^*$ is antichiral $\de_{\a i}\bar{U}=0$.
In writing these results, we have introduced second and fourth order spinorial derivative
operators via the equations 
 \be
 \eqalign{ 
~~&\nabla{}^{(2)}_{\a \, \b}  ~=~  \fracm 12 \, C^{i \, j} \left[ \,
\nabla_{\a \, i} \, \nabla_{\b \, j}  ~+~ \nabla_{\b \, i} \, \nabla_{\a \, j} 
\, \right] ~, ~   
\nabla{}^{(2)}_{i \, j}   ~=~  \fracm 12 \, C^{\a \, \b} \left[ \,
\nabla_{\a \, i} \, \nabla_{\b \, j}  ~+~ \nabla_{\a \, j} \, \nabla_{\b \, i} 
\, \right]  ~, \cr 
%%%%%%%%%%%%%%%%%%%%%%%%%%%%%%%%%%%%%%%%%%%%%
~~&{\Bar \nabla}{}^{(2)}_{\a \, \b}  ~=~  \fracm 12 \, C_{i \, j} \left[ \,
{\Bar \nabla}_{\a} {}^{ i} \, {\Bar \nabla}_{\b}{}^{j}  ~+~ {\Bar \nabla}_{\b}{}^{i} \, {\Bar \nabla}_{\a}{}^{j} 
\, \right] ~, ~   
{\Bar \nabla}{}^{(2)}{}^{i \, j}   ~=~  \fracm 12 \, C^{\a \, \b} \left[ \,
{\Bar \nabla}_{\a} {}^{ i} \, {\Bar \nabla}_{\b}{}^{j}  ~+~ {\Bar \nabla}_{\a}{}^{j} \, {\Bar \nabla}_{\b}{}^{i} 
\, \right]  ~, \cr 
%%%%%%%%%%%%%%%%%%%%%%%%%%%%%%%%%%%%%%%%%%%%%
~~&  \nabla{}^{(4)}   ~=~ \fracm 13\nabla{}^{(2)}{}^{k \, l}  \, \nabla{}^{(2)}_{k \, l}  ~~~,~~~~ 
{\Bar  \nabla}{}^{(4)}   ~=~ \fracm 13 {\Bar \nabla}{}^{(2)}{}^{k \, l}  \, {\Bar \nabla}{}^{(2)}_{k \, l} 
~~~.}  \label{DiffOps}  \ee

The proof of the theorem involves using the above equations to show that the Bianchi identities 
for this two-form vanish. This is relegated to an appendix.  We next note that the chiral 
superfield $U$ above may be replaced using the result from (\ref{ChRLcond}) according to
$U=\bar{\cal D}^{(4)}{\cal L}$ ($\bar{U}={\cal D}^{(4)}\bar{\cal L}=(\bar{\cal D}^{(4)}{\cal L})^*$)
where the 2D, $\cal N$ = 4 superfield $\cal L$,  ($\bar{\cal L}:=({\cal L})^*$),
is not subject to any algebraic nor
differential restrictions.  Stated another way, this implies
that an arbitrary 2D, $\cal N$ = 4 superfield  $\cal L$ can be used to create a closed 
super 2-form whose components are defined by $J_{A \, B}$ above.  
We conclude with a result that will be needed in the next section.
Defining the component vierbein
$E_m{}^a|=e_m{}^a$ ($e_m{}^ae_a{}^n=\d_m^n$, $e_a{}^me_m{}^b=\d_a^b$), 
and the gravitini $E_m{}^{\a i}|=-\psi_m{}^{\a i}$ ($\psi_a{}^{\a i}=e_a{}^m\psi_m{}^{\a i}$), 
$E_m{}^{\a}_i|=-\bar{\psi}_m{}^{\a}_{i}$ ($\bar{\psi}_a{}^{\a}_{i}=e_a{}^m\bar{\psi}_m{}^{\a}_{i}$), 
by a general result given in \cite{ggrs1,Ecto} taking the limit as all Grassmann coordinates
go to zero one obtains
\be
\begin{array}{lll} 
~~  \ve^{ab} {J}_{a b } {\Big |} &=&  \ve^{ab} {\cal J}_{a b } {\Big |} ~+~ 2
\,  \ve^{ab} (\psi_{ a} 
{}^{\a i} {J}_{\a i ~b  }  {\Big |}~+~ {\bar \psi}_{a} {}^{ \a}{}_i {J}_{ \a}{}^i{}_{ ~b }  {\Big |})
~+~ 2 \,   \ve^{ab} \psi_{ a} {}^{\a i}{\bar \psi}_{ b} {}^{ \b}{}_j 
{J}_{\a}{}_i {}_{ \b}{}^j {\Big |}  \\ 
& &    \\
&{\,}& +~   \ve^{ab} \psi_{a} {}^{\a i } \psi_{ b}{}^{\b j}  {J}_{\a \, i \, \b \, j } {\Big |} 
~+~  \ve^{ab} {\bar \psi}_{a} {}^{ \a} {}_i
{\bar \psi}_{b} {}^{ \b}{}_j {J}_{\a}{}^i{}_{  \b  }{}^j  {\Big |}
~~~, \end{array}           \label{FormStrength}
\ee
where ${\cal J}_{a b } {\Big |}$ describe an ordinary space closed 2-form.

 %%%%%%%%%%%%%%%%%%%%%%%%%%%%%%%%%%%%%%%%%%
\section{A 2D, ${\cal N} = 4$ Density Projection Operator }

~~~~ It remains for us to calculate
 the explicit form of the density projection operator 
(that we will denote by ${\D}^{(4)}$) which is the main
purpose of this work. As we are going to describe in this section,
by using ${\D}^{(4)}$ and the chiral projector $\overline{{\cal D}}^{(4)}$,
we can build the integration measure of component actions in 2D, ${\cal N}=4$ 
minimal supergravity.

As noted by Siegel \cite{Fields}, the `secret' to the ectoplasmic
approach is to realize that the integration theory of superspace can be totally cast into
the language of closed super-forms. Indeed it was argued in the work of \cite{Ecto} that
the requirement that the topology of a superspace be totally determined by the topology
of its purely bosonic sub-manifold 
naturally provides a reason for the appearance
of super-forms in constructing integration measures of superspace.

 In the work of Ref.~\cite{ggrs1} it was noted that the derivation of
component results follows 
 efficiently from replacing the integration
of fermionic coordinates by a process using {\em {first}} application of the 
superspace covariant derivative followed by taking the limit as the Grassmann
coordinates are taken to zero.  In the 2D, $\cal N$=4 case, this is described in the form of an equation
\be \eqalign{
\int d^2 \s\, d^4 \theta \,  d^4 {\Bar \theta\ } {\rm E}^{-1}~{\cal L} &\to ~ \int d^2 \s\ 
\, \fracm 12 \,  {\rm e}{}^{-1}  \, {\Big [} \,   {\D}{}^{(4)}  \, { {\cal {\Bar D}}{}^{(4)} }  \,
\,  {\cal L} ~+~ {\rm h}. \, {\rm c}. \, {\Big ]} \, {\Big |}
  }    \label{Sint} \ee 
in terms of two differential operators, ${ {\D}}^{(4)}$ (the density 
projection operator) and $ {\cal {\Bar D}}{}^{(4)}$ (the chiral projection operator)
which may be expanded as
\be
{ {\D}}{}^{(4)} ~=~ \sum_{i = 0}^{4} \, b_{( 4 - i)} \,  \cdot\,  \, \left[ \,
(\nabla) \, \times\,  \, \cdots\, \, \times\,( \nabla)^{4 - i}  \, \right]
~~~,  \label{exp1} 
 \ee
 \be
 {\cal {\Bar D}}{}^{(4)} ~=~ \sum_{i = 0}^{4} \, a_{( 4 - i)} \, 
\cdot\,  \, \left[ \, (\Bar \nabla) \, \times \, \cdots\,  \, \times
\,({\Bar  \nabla})^{4 - i}  \, \right]
~~~,
 \label{exp2} 
 \ee
in terms of some field-dependent coefficients $a_{(4 - i)}$ and $b_{(4 - i)}$ and powers 
of the spinorial superspace supergravity covariant derivatives $\nabla_{\a \, i}$ and 
${\Bar \nabla}{}_{\a}{}^{ i}$.  
In (\ref{Sint}) we have the expressions ${\rm E}^{-1}=[{\rm Ber}\,{E_{A}{}^M}]^{-1}$ and
${\rm e}^{-1}=[\det{e_{a}{}^m}]^{-1}$ which are functions respectively of the supervielbein and the 
component vielbein and $d^2 \s$ denotes the measure over the two-dimensional bosonic space.
A further consequence of (\ref{Sint}) - (\ref{exp2}) 
is that the superfield Lagrangian $\cal L$ need not be hermitian as it is the linear
combination of terms that appear in the action formula that must satisfy this requirement.
In the present context these spinorial superspace 
supergravity covariant derivatives satisfy the relations given in section 2.

The basis for the ectoplasmic derivations of local supergravity measures and projections
operators, lies in a proposition for how to integrate an arbitrary super $p$-form.  This was proposed
in the work of \cite{Ecto}. 
Given a curved superspace with ${\rm N}_B$ bosonic coordinates 
(labelled by $\underline{m}$ indeces) 
and ${\rm N}_F$ fermionic coordinates
(labelled by $\underline{\mu}$ indices),
we have
\vskip.05in 
\noindent
$~~$ {\it {Proposition 1}} \newline  \noindent
$~~~$ {\it {If $J{}_{{\underline A}_1 \dots {\underline A}_p}$} is a closed super $p$-form
superfield whose Bianchi identities}   \newline  \noindent
$~~~$ {\it {vanish and $d\O{}^{{\underline m}_1 \dots {\underline m}_p}$ is a co-chain of
dimension $p$ $\le$ {\rm N}${}_B$} (where {\rm N}${}_B$ is}
   \newline  \noindent
$~~~$ {\it {the dimensionality of the
bosonic subspace), then the integral of the su- }
 \newline  \noindent
 $~~~$ {\it {per $p$-form over the co-chain is given by}
\be \eqalign{
{\cal S} (d\O{\, |} J) ~&\equiv ~ (p!)^{-1} \, \int ~ d \O^{{\un a}_1 
\cdots {\un a}_{p} } \, {\cal J}^{(p)}_{{\un a}_1 \cdots {\un a}_p }  {\Big |}~~~.
}  \label{IntTheory}  \ee
}} {\it {$~$} and this is a supersymmetrical invariant. }  \vskip.1in
\noindent
In (\ref{IntTheory}) we note the quantity ${\cal J}^{(p)}_{{\un a}_1 \cdots {\un a}_p } {\Big |}$ is related
to the super $p$-form $J{}_{{\underline A}_1 \dots {\underline A}_p}$ via
\begin{equation}
\eqalign{
 \Big( \, J_{{\un a}_1 \cdots {\un a}_{p} } {\Big |} \, \Big)
 ~\equiv~  \Big[ & ~
{\cal J}^{(p)}_{{\un a}_1 \cdots {\un a}_p } {\Big |} \,+\, \l^{(p , 1)} \psi_{ [ \, {\un a}_1 | }{}^{{\un \a}_1} 
\Big( \, J_{{\un \a}_1  | \, {\un a}_2  \cdots {\un a}_{p} \, ] } {\Big |} \, 
\Big)   \cr 
&+\, \l^{(p , 2)} \psi_{ [\, {\un a}_1 |}{}^{{\un \a}_1} \psi_{ |{\un a}_2
| }{}^{{\un \a}_2} \Big( \, J_{ {\un \a}_1 {\un \a}_2  | \,{\un a}_3 \cdots {\un a}_{p} 
\, ] }  {\Big |} \, \Big) \cdots \cr
&+\, \l^{(p , p)}\, [\,\psi_{ {\un a}_1}{}^{{\un \a}_1} \cdots \psi_{ 
{\un a}_{p}}{}^{{\un \a}_{p}} \,] \Big( \, J_{{\un \a}_1 {\un \a}_2 \cdots {\un \a}_{p} } {\Big |} 
\, \Big) ~\Big] ~~~.
 }   \label{IntTheory2}
\end{equation}
where $ \psi_{ {\un a} }{}^{{\un \a}} $ denotes the gravitino.  
The quantities ${\cal J}^{(p)}_{{\un a}_1 \cdots 
{\un a}_p } {\Big |}$ and coefficients $\l^{(p , 1)}\,$ $\cdots$ $\l^{(p , p)}\,$ are determined by 
taking the limit as the Grassmann coordinates go to zero in $J_{{\un a}_1 \cdots {\un a}_{p} } $.  In the 2D, ${\cal N}=4$ case with $J_{ab}$ the component of a super 2-form,
 the equation (\ref{FormStrength}) informs us about the $\l$-coefficients.''

We next observe that upon setting $p$ = N${}_B$ the proposition takes the form
\be \eqalign{
{\cal S} (d\O{\, |} J) ~
&=~ \int ~ d^{N_B}x \, {\rm e}^{-1} \,   \fracm 1{~{\rm N}{}_B! ~} \,  \ve^{{\un a}_1
 \cdots {\un a}_{N_B}} \, {\cal J}^{({\rm N}{}_B)}_{{\un a}_1 \cdots {\un a}_{{\rm N}{}_B} }  {\Big |}
~~~,
}  
\ee
where e${}^{-1}$ denotes the determinant of the vielbein for the bosonic subspace.
In the case considered in this paper, we thus reach the result
\be \eqalign{ {~~~~~~}
{\cal S} (d\O{\, |} J) ~
&=~ \int ~ d^{2} \s  \, {\rm e}^{-1} \,   \fracm 1{2} \,  \ve^{a \, b}
\, {\cal J}^{(2)}_{ a \, b }  {\Big |}  \cr
&=~ \int ~ d^2 \s  \, {\rm e}^{-1} \, {\Big [} ~ \fracm 1{2} \,  \ve^{a \, b}
\, {J}_{ a \, b } {\Big |} ~-~   \ve^{ab} (\psi_{ a} 
{}^{\a i} {J}_{\a i ~b  }  {\Big |}~+~ {\bar \psi}_{a} {}^{ \a}{}_i {J}_{ \a}{}^i{}_{ ~b }  {\Big |})\cr
&~~~~~~~~~~~~~~~~~~~~~-~  \,   \ve^{ab} \psi_{ a} {}^{\a i}{\bar \psi}_{ b} {}^{ \b}{}_j 
{J}_{\a}{}_i {}_{ \b}{}^j {\Big |}   \cr
&~~~~~~~~~~~~~~~~~~~~~-~ \frac 12 \,  \ve^{ab} \psi_{a} {}^{\a i } \psi_{ b}
{}^{\b j}  {J}_{\a \, i \, \b \, j } {\Big |} ~-~  \frac12 \,  \ve^{ab} {\bar \psi}_{a} {}^{ \a} {}_i
{\bar \psi}_{b} {}^{ \b}{}_j {J}_{\a}{}^i{}_{  \b  }{}^j  {\Big |} ~{\Big ]}
~~~.
}  \label{IntTheory4}  \ee
More explicitly the equations in (\ref{2Form}) are expressed as
\be \eqalign{
 {~~}
 J_{\a i}{}_{\b}{}^{ j} &=~ 0 
~~~,  \cr
%%%%%%%%%%%%%%%%%%%%%%%%%%%%%%%%%%
 J_{\a i}{}_{\b j} &=~
2(\g^3)_{\a\b}\deb^{(2)}_{ij} \left[ \, { {\nabla}}{}^{(2) \, \e \, \k}  ~-~ 2 \, 
{\bar  B} \, (\g^3){}^{\e \, \k} \, \right] \, { {\nabla}}{}^{(2)}_{ \, \e \,  \k} \, 
{\Bar {\cal L}}  \cr
%%%%%%%%%%%%%%%%%%%%%%%%%%%%%%%%%%
&~~~~~
-\,C_{\a\b}C_{ij}(\g^3)^{\g\d}\deb^{(2)}_{\g\d} \left[ \, { {\nabla}}{}^{(2) \, 
\e \, \k}  ~-~ 2 \, {\bar  B} \,  (\g^3){}^{\e \, \k} \, \right] \, { {\nabla}}{}^{(2)}_{ 
\, \e \, \k} \,{\Bar {\cal L}}
~~~,  \cr
%%%%%%%%%%%%%%%%%%%%%%%%%%%%%%%%%%
 J_{\a}{}^i{}_{\b}{}^j &=~
 \, 2 \, (\g^3)_{\a\b}\de^{(2) \,ij} \left[ \, {\Bar {\nabla}}{}^{(2) \, \e \, \k}  ~-~ 
2 \, B \, (\g^3){}^{\e \, \k} \, \right] \, {\Bar {\nabla}}{}^{(2)}_{ \, \e \, \k} \,{\cal 
L}   \cr
%%%%%%%%%%%%%%%%%%%%%%%%%%%%%%%%%%
&~~~~~
~-~ \,C_{\a\b}C^{ij}(\g^3)^{\g\d}\de^{(2)}_{\g\d}
\left[ \, {\Bar {\nabla}}{}^{(2) \, \e \, \k}  ~-~ 2 \, B \, (\g^3){}^{\e \, \k} \, \right] \, 
{\Bar {\nabla}}{}^{(2)}_{ \, \e \, \k} \,{\cal L}
~~~,
  \cr
%%%%%%%%%%%%%%%%%%%%%%%%%%%%%%%%%%
J_{\g k}{}_a &=~ -~
\fracm \ri3\ve_{ab}(\g^b)_{\g}{}^{\d}\deb_\d{}^{p}\deb^{(2)}_{kp}
\left[ \, { {\nabla}}{}^{(2) \, \e \, \k}  ~-~ 2 \, {\bar  B} \, 
(\g^3){}^{\e \, \k} \, \right] \, { {\nabla}}{}^{(2)}_{ \, \e \, 
\k} \,{\Bar {\cal L}}
~~~,
 \cr
%%%%%%%%%%%%%%%%%%%%%%%%%%%%%%%%%%
J_{\g}{}^k{}_a
&=~ -~
\fracm \ri3\ve_{ab}(\g^b)_{\g}{}^{\d}\de_{\d p}\de^{(2) \,kp}
 \left[ \, {\Bar {\nabla}}{}^{(2) \, \e \, \k} ~-~ 2 \, B \, (\g^3){}^{\e \, \k} 
 \, \right] \, {\Bar {\nabla}}{}^{(2)}_{ \, \e \, \k} \,{\cal L}
~~~,  \cr
%%%%%%%%%%%%%%%%%%%%%%%%%%%%%%%%%%
J_{ab}
&=~
-~\fracm 18 \,  \ve_{ab}\,  \Big[ \de^{(4)} ~-~ 2 \, \bar{B}(\g^3)^{\a\b}
\de^{(2)}_{\a\b}\Big]  \left[ \, {\Bar {\nabla}}{}^{(2) \, \e \, \k}  ~-~ 2 \, B \, 
(\g^3){}^{\e \, \k} \, \right] \, {\Bar {\nabla}}{}^{(2)}_{ \, \e \, \k} \,{\cal L}
\cr
%%%%%%%%%%%%%%%%%%%%%%%%%%%%%%%%%%
&~~~~~
-~
\fracm 18 \, \ve_{ab} \, \Big[ ~ \deb^{(4)} ~-~ 2 \, B(\g^3)^{\a\b} \deb^{
(2)}_{\a\b}\Big] \left[ \, { {\nabla}}{}^{(2) \, \e \, \k}  ~-~ 2 \, {\bar  B} \, (\g^3)
{}^{\e \, \k} \, \right] \, { {\nabla}}{}^{(2)}_{ \, \e \, \k} \,{\Bar {\cal L}}
~~~.
  } \label{2FormnoU2}   \ee

Finally, the results in (\ref{2FormnoU2}) can be substituted into equation
(\ref{IntTheory4}) to reach the main result of this presentation.  Given an
arbitrary 2D, $\cal N$ = 4 superfield Lagrangian $\cal L$, a local supersymmetrical
invariant is given by
\be \eqalign{ 
{\cal S}  &=~   \int ~ d^{2} \s \, {\rm e}^{-1} \,  \D^{(4)} \,\bar{\cal D}^{(4)}\,{\cal L}  {\Big |} 
~+~ {\rm h. ~c.}
\cr
&=~ \int ~ d^{2} \s \, {\rm e}^{-1} \, {\Big \{} ~
  \fracm 18 \de^{(4)} ~-~  \fracm 14 \, \bar{B}(\g^3)^{\a\b}\de^{(2)}_{\a\b}
~+~   \fracm \ri 3{\bar \psi}_{a} {}^{ \g}{}_i ( \g^a)_{\g}{}^{\d}\de_{\d j}\de^{(2) \,i j}
\cr &~~~~~~~~~~~~~~~~~~~~~
-~    \ve^{ab} {\bar \psi}_{a} {}^{ \a} {}_i {\bar \psi}_{b} {}^{ \b}{}_j (\g^3)_{\a\b}\de^{(2) \,ij} 
~+~  \fracm 12\,  \ve^{ab} {\bar \psi}_{a} {}^{ \a} {}_i
{\bar \psi}_{b} {}^{ \b}{}_j C_{\a\b}C^{ij}(\g^3)^{\g\d}\de^{(2)}_{\g\d}
~{\Big \} } \times
\cr &~~~~~~~~~~~~~~~~~~~~
\times\left[ \, {\Bar {\nabla}}{}^{(2) \, \e \, \k}  ~-~ 2 \, B \, (\g^3){}^{\e \, \k} \, \right] \, 
{\Bar {\nabla}}{}^{(2)}_{ \, \e \, \k} \,{\cal L}  {\Big |} 
~+~ {\rm h. ~c.}
}  \label{IntTheory5}  \ee
in the presence of the off-shell supergravity theory described in section 2.

%%%%%%%%%%%%%%%%%%%%%%%%%%%%%%%%%%%%%%%%%%
\section{Conclusion }

~~~~ With this present work, we have completed the task of developing an efficient
local superspace integration theory for two dimensional theories that possess eight real
supercharges.  We believe that the result given in (\ref{IntTheory5}) is
unexpectedly elegant and simple given that the general form of the eigth-order
spinorial differential operator defined by (\ref{Sint}), (\ref{exp1}) and (\ref{exp2})
could, in principle, take a more complicated form.  Perhaps one of most
surprising features of this derivation has been the use of the closed 2D, $\cal N$
= 4 super 2-form used in Theorem 1.  The superfield $U$ that appeared in
equation (\ref{2Form}) is {\em {not}} required to describe any irreducible supermultiplet.  
The only requirement imposed on the superfield $U$ is its chirality.  

As proved in \cite{GM}, the chiral superfield $U$ can be expressed in terms of the 
chiral projector $\overline{\cal{D}}^{(4)}$ and an unconstrained superfield ${\cal {L}}$ as 
$U=\overline{\cal{D}}^{(4)}{\cal{L}}$. This result has been used in sections 3 and 4.
According to the discussion of section 4, the main result of this paper is the computation 
of the density projector operator $\D^{(4)}$ of (\ref{Sint}), (\ref{exp1}) and (\ref{IntTheory5}),
which, together with $\overline{\cal{D}}^{(4)}$, allows to define the component supergravity 
integration measure  (\ref{IntTheory5}). In deriving for the first time $\D^{(4)}$ 
we used the Ectoplasmic techniques
and the new super 2-form of Theorem 1 (\ref{2Form}).

One other point we wish to emphasize is the efficiency of the Ectoplasmic approach
in the case we considered here. It would be interesting to re-derive  the integration 
measure (\ref{IntTheory5}) via the normal coordinate expansion technique 
\cite{Norcor,Ectonorcor,KuzGtmNormCoord} (in particular using its last version 
\cite{KuzGtmNormCoord})
even if we do not expect that the latter approach
would require shorter computations.  This is especially  true considering that in 2D 
the number of Bianchi identities to be solved for a closed super 2-form is relatively 
low.  This once more emphasizes the important role of forms as a basis for superspace 
integration theory as advocated in the ectoplasmic approach.  The success of this 
also points to the generality of using this as a tool in all cases to derive superspace
local integration measures.

Note also that here we focused on the 2D, ${\cal N}=4$ 
minimal superspace geometry of \cite{2DN4SG} as described in section 2.
In general, it is known that there could exist
different off-shell superspace supergravities.
We expect that the Ectoplasm paradigm and the results of our paper can be extended to 
any covariant superspace formulation of  2D, ${\cal N}=4$ supergravity.
For example, in the first paper of \cite{2DN4SG} a variant central charge formulation of the minimal 
multiplet was given; once noticed that the Lagrangian ${\cal L}$ in (\ref{IntTheory5}) has to be 
neutral for the central charges, one can see that the results of our paper apply without 
modifications to the variant formulation.
Moreover, recently a new extended covariant formulation of 2D, ${\cal N}=4$ supergravity in 
superspace was given \cite{GTM-2DSG}. The ectoplasm techniques to compute the chiral
action in components apply straightforward if one consider the geometry of \cite{GTM-2DSG}
even if in this case 
longer computations are expected
 due to  the more involved structure of the torsion multiplet. 
Other superspace formulations of 2D, ${\cal N}=4$ supergravity \cite{BellucciIvanov}
are known in the bi-harmonic superspace of \cite{IvanovSutulin}.
Being those superspace supergravities based on a prepotential approach the definition of a 
covariant components reduction is not  clear. However, on the ground 
of the related bi-projective formalism \cite{biProj}, recently extended to covariantly study
2D, ${\cal N}=4$ matter-couplet supergravity, it would be of interest and well defined 
 to find by using Ectoplasm techniques the bi-projective density 
operator analogously to the chiral action studied here.

 \vspace{.1in}
 \begin{center}
 \parbox{4in}{{\it ``Where the senses fail us, reason must step in.''}\newline$~~$
   \,\,-\,\, Galileo Galilei}
 \end{center}

 \vspace{.2in}

 \noindent

 {\bf Acknowledgements}\\[.1in] \indent
This research was supported in part by the endowment of the John S.~Toll Professorship,
the University of Maryland Center for String \& Particle Theory, National Science Foundation 
Grant PHY-0354401.

\newpage
\noindent
{\Large{\bf Appendix A:  Consistency of Bianchi Identities \& 
$~~~~~~~~~~~~~~~~~\,~$ Constraints For 2-Form}}
\setcounter{equation}{0}

~~~~ In this appendix, we will present the explicit proof that the Bianchi identities associated with 
the results in (\ref{2Form}) imply that it is a closed 2D, $\cal N$ = 4 super 2-form.  We begin
by writing an ansatz
for the lowest components of a 
2D, $\cal N$ = 4 super 2-form under the assumption that these component should:
\newline \indent
(a.)  be linear in a (anti)chiral superfield $U$ $(\bar{U})$; $\deb_{\a}{}^{i}U=0$ 
$(\de_{\a i}\bar{U}=0)$,
\newline \indent
(b.) depend on the superspace supergravity covariant derivative,
\newline \indent
(c.) and are local functions of the superspace supergravity field strengths
\newline \indent $~~~~$
$B$, $\Bar B$, $G$ and $H.$ 
\newline \noindent
Under the previous assumptions we will begin with an ansatz
 given by
\footnote{The ansatz we are using can also  be guessed by: 
(i) consider the flat 4D, $\cal N$=2 ``chiral'' closed \\ $~~~$$~~~$
 super 4-form introduced in \cite{BiswasSiegel};
(ii) perform a dimensional reduction of the 4D, $\cal N$=2 super \\ $~~~$$~~~$
 4-form to derive a 
2D, $\cal N$=4 closed super 2-form;
(iii) extend the resulting dimension-1 \\ $~~~$$~~~$
 components of the flat 2D, $\cal N$=4  2-form to the curved case by modifying the 
flat derivatives to \\ $~~~$$~~~$
 the curved covariant derivatives and by adding torsion dependent terms.} 
$$ \eqalign{
J_{\a i}{}_{\b j}=
a(\g^3)_{\a\b}\deb^{(2)}_{ij}\bar{U}
+b\,C_{\a\b}C_{ij}(\g^3)^{\g\d}\deb^{(2)}_{\g\d}\bar{U}
+C_{\a\b}C_{ij}F\bar{U}~,~~~
} \eqno({\rm A}.1) $$
$$ \eqalign{
{J}_{\a i}{}_\b{}^j=0~,~~~~~~
{J}_\a{}^i{}_\b{}^j=-(J_{\a i}{}_{\b j})^*~,
} \eqno({\rm A}.2) $$
where
$$ \eqalign{
F=F(B,\bar{B},G,H)=b_1B+b_2\bar{B}+gG+hH~,
} \eqno({\rm A}.3) $$
and $a,\, b,\, b_1,\,b_2,\,g,\,h$ are constants to be fixed.

The task is to
study the Bianchi identities that derive from the closure of the 2-form $J$
$$
d J=0~,~~~~~~\Longleftrightarrow~~~~~~
0={1\over 2}\de_{[A}J_{BC)}-{1\over 2}T_{[AB|}{}^DJ_{D|C)}~,
\label{C-dJ=0}
\eqno({\rm A}.4) $$
with $  J_{A}
{}_{B} =\big(\,  J_{\a i}{}_{\b j} , ~  J_{\a  i}{}_{\b}{}^{j} ,~  J_{\a}{}^i{}_{\b}{}^j , ~ 
J_{\g k}{}_{a}, ~ J_{\g}{}^k{}_{a} , ~ J_{ab}\, \big)$ and the lowest components satisfying 
(A.1) - (A.3).

Substituting the results of (A.1), (A.2) into the identity (A.4)  with $A=\a i,\,B=\b j,\,C=\g k$
one obtains
$$ \eqalign{
{~~~~~~~} 0&=~ 
a(\g^3)_{\b\g}[\de_{\a i},\deb^{(2)}_{jk}]\Bar{U}
+a(\g^3)_{\g\a}[\de_{\b j},\deb^{(2)}_{ki}]\Bar{U}
+a(\g^3)_{\a\b}[\de_{\g k},\deb^{(2)}_{ij}]\Bar{U}
\cr
&~~~~
+b\,C_{\b\g}C_{jk}(\g^3)^{\d\rho}[\de_{\a i},\deb^{(2)}_{\d\rho}]\Bar{U}
+b\,C_{\g\a}C_{ki}(\g^3)^{\d\rho}[\de_{\b j},\deb^{(2)}_{\d\rho}]\Bar{U}
\cr
&~~~~
+b\,C_{\a\b}C_{ij}(\g^3)^{\d\rho}[\de_{\g k},\deb^{(2)}_{\d\rho}]\Bar{U}
+C_{\b\g}C_{jk}\, (\de_{\a i}F) \Bar{U}
+C_{\g\a}C_{ki} \, (\de_{\b j}F) \Bar{U}\cr
&~~~~
+C_{\a\b}C_{ij}\, (\de_{\g k}F)\Bar{U}~,
~~~~~~
\label{C-3/2-1-1}
} \eqno({\rm A}.5) $$
where we have used the fact that $\bar{U}$ is antichiral to write this.
At this point, there are two useful identities to note
$$
{[}\de_{\a i}, \deb^{(2)}_{i j}{]}\Bar{U}  ~=~
\Big(-2\ri C_{i(j}(\g^c)_{\a}{}^{\d}\de_c\deb_{\d k)}
\Big) \Bar{U}
\eqno({\rm A}.6)  $$
$$
{[}\de_{\a i},(\g^3)^{\d\rho} \deb^{(2)}_{\d\rho}{]}  ~=~
\Big(
- 4\ri \ve^{bc}(\g_b)_{\a}{}^{\b}\de_c\deb_{\b i}
\Big)
\Bar{U}
\eqno({\rm A}.7) $$
which shows that in principle there are terms containing spacetime derivatives in (A.5).  
In order to satisfy the Bianchi identity, two sets of conditions are required:
$$  \eqalign{
{~~} &(a.)~~  a ~=~-2b ~~{\rm {and}}  \cr 
 &(b.)~~ b_1~=~ b_2 ~=~g ~=~ h ~= ~0  ~~. 
} \eqno({\rm A}.8)
$$
For simplicity we also set
$$
a=1~.\eqno({\rm A}.9)
$$

The next Bianchi identity encountered takes the form
$$ \eqalign{
0 &=~
\deb_\a^iJ_{\b j}{}_{\g k}
+T_\a{}^i{}_{\b j}{}^aJ_{\g k}{}_a
+T_\a{}^i{}_{\g k}{}^aJ_{\b j}{}_{a}
~~~.
} \eqno({\rm A}.10) $$
The result in (A.1), subject to (A.8), (A.9), can be substituted into this equation.  To
satisfy this, it is useful to use the following identities
$$  \eqalign{
\deb_{\a i}\deb^{(2)}_{jk}\bar{U} &=~- \, \fracm 13
C_{i(j}\deb_\a^{p}\deb^{(2)}_{k)p} \Bar{U}
~~,  \cr
\deb_{\a i}\deb^{(2)}_{\b\g}\bar{U}&=~ \fracm 13
C_{\a(\b}\deb_{\g)}^p\deb^{(2)}_{ip}\Bar{U}
~-~ \fracm 43 \, BC_{\a(\b}(\g^3)_{\g)}{}^\d \deb_{\d i}\Bar{U}
~+~ \fracm 13 \, B(\g^3)_{(\a\b}\deb_{\g)i}\Bar{U}  ~~, \cr
(\g^3)^{\b\g}\deb_{\a i}\deb^{(2)}_{\b\g}\bar{U}&=~
- \, \fracm 23 \, (\g^3)_\a{}^{\g}\deb_{\g}^p\deb^{(2)}_{ip}\Bar{U}  ~~~.
}  \eqno({\rm A}.11)$$
Then, to completely satisfy (A.10)
one has to impose
$$
J{}_{\g k}{}_{a}
~=~ - ~\fracm \ri3 \,
\ve_{ab}(\g^b)_{\g}{}^{\rho}\deb_\rho{}^{p}\deb^{(2)}_{kp}\bar{U}
  ~~~.
\eqno({\rm A}.12) $$
Note that it holds
$$
J{}_{\g}{}^{k}{}_{a}~=~
-(J{}_{\g k}{}_{a})^*
~=~ - ~\fracm \ri3 \,
\ve_{ab}(\g^b)_{\g}{}^{\rho}\de_{\rho p}\de^{(2)}{}^{kp}{U}
  ~~~.
\eqno({\rm A}.13) $$

We can continue our deliberations by considering the Bianchi identity given by
$$
0 ~=~
\de_{a}J_{\b j \g k}
~+~ \de_{\b j}J_{\g k}{}_a
~+~ \de_{\g k}J_{\b j}{}_a
~-~ T_{a\b j}{}^{\d l}J_{\d l \g k}
~-~ T_{a\g k}{}^{\d l}J_{\d l \b j}
~~~,
\eqno({\rm A}.14)  $$
and into this are substituted the results 
(A.1), (A.8), (A.9) and (A.12).  When this
is done, a differential equation on $\Bar U$ of the form
$$ \eqalign{  {~~~~~~~}
0 ~=&~ \de_{a}\Big( 2(\g^3)_{\b\g}\deb^{(2)}_{jk}
-\,C_{\b\g}C_{jk}(\g^3)^{\d\rho}\deb^{(2)}_{\d\rho} \Big)\Bar{U}  \cr
&~ -~ \fracm \ri3\,  \de_{\b j}\Big(\ve_{ab}(\g^b)_{\g}{}^{\d}\deb_\d{}^{p}\deb^{(2)}_{kp}\Bar{U}
\Big) ~-~ \fracm \ri3\,  \de_{\g k}\Big(\ve_{ab}(\g^b)_{\b}{}^{\d}\deb_\d{}^{p}\deb^{(2)}_{jp}\Bar{U}
{\Big)}
\cr
&
~+~ \fracm \ri2 \, \d_j^l\phi_\b{}^\rho(\g_a)_\rho{}^\d \Big(2(\g^3)_{\d\g}\deb^{(2)}_{lk} 
-\,C_{\d\g}C_{lk}(\g^3)^{\rho\tau}\deb^{(2)}_{\rho\tau}\Big)\Bar{U}   \cr
&
~+~ \fracm \ri2\,  \d_k^l\phi_\g{}^\rho(\g_a)_\rho{}^\d \Big(2(\g^3)_{\d\b}\deb^{(2)}_{lj}
-\,C_{\d\b}C_{lj}(\g^3)^{\rho\tau}\deb^{(2)}_{\rho\tau}\Big)\Bar{U}
} \eqno({\rm A}.15) $$
emerges.  Further progress is possible by using the identity 
$$  \eqalign{  {~~}
\{\de_{\a i},\deb_\d^{p}\deb^{(2)}_{kp} \}\Bar{U} &=~
 \Bigg(
3\ri(\g^a)_{\a\d}\de_a\deb^{(2)}_{ik}
-3\ri C_{ik}(\g^a)_{\a}{}^{\rho}\de_a\deb^{(2)}_{\d\rho} \cr
& {~~~~~~~~~} 
+~ \fracm 32 C_{ik}{\phi}^\tau{}_\d(\g^3)_{\a\tau}(\g^3)^{\rho\b}\deb^{(2)}_{\b\rho}
-3\phi_{\a\d}\deb^{(2)}_{ik}
\cr
& {~~~~~~~~}
~+~6C_{ik}C_{\a\d}\S^{\b p}\deb_{\b p}
~+~ 6C_{ik}(\g^3)_{\a\d}(\g^3)^{\b\rho}\S_{\b}{}^p\deb_{\rho p}
\Bigg)\Bar{U}
~~.
}  \eqno({\rm A}.16)
$$
This result is substituted into (A.15) and after some algebra, the $\S$-dependent
terms are seen to cancel leaving
$$  \eqalign{  {~~~~~~~~~~~}
0 \, &=
\Bigg(
2(\g^3)_{\a\g}\de_{a}\deb^{(2)}_{ik}
~-~ C_{\a\g}C_{ik}(\g^3)^{\d\rho}\de_{a}\deb^{(2)}_{\d\rho}  \cr
&~~~~~~~
-2(\g^3)_{\a\g}\de_a\deb^{(2)}_{ik}
+C_{ik}C_{\a\g}(\g^3)^{\d\rho}\de_a\deb^{(2)}_{\d\rho}
\cr
&~~~~~~~
-~\ri\ve_{ab}\phi_{\b\d}(\g^b)_{\a\g}C^{\b\d}\deb^{(2)}_{ik}
~-~\ri\phi_{\b\d}(\g^3)^{\b\d}(\g_a)_{\a\g}\deb^{(2)}_{ik}
\cr
&~~~~~~~
+~ \ri\ve_{ab}\phi_{\a'\d}(\g^b)_{\a\d}C^{\a'\d}\deb^{(2)}_{ik}
~+~ \ri\phi_{\a'\d}(\g_c)_{\a\d}(\g^3)^{\a'\d}\deb^{(2)}_{ik}
\Bigg)\Bar{U}
~~
}  \eqno({\rm A}.17)
$$
which is clearly identically satisfied.

There is a second dimension-2 Bianchi identity of the form
$$  \eqalign{  {~~~~~~~~~~}
0 \, &=~-~\deb_\a{}^iJ_{\g k}{}_b
~-~\de_{\g k}J_\a{}^i{}_b
~+~T_b{}_{\g k}{}^\d{}_lJ_\d{}^l{}_\a{}^i
~+~T_b{}_\a{}^i{}^{\d l}J_{\d l}{}_{\g k}
~+~T_{\g k}{}_\a{}^i{}^cJ_{cb}   ~~.
}  \eqno({\rm A}.18)
$$
One may substitute from results derived previously to cast this into the form
of
$$ \eqalign{
2\ri\d^i_k(\g^c)_{\a\g}J_{bc}
&=~\ri  \Big( \fracm 13\ve_{bc}(\g^c)_{\g}{}^{\rho}\deb_\a{}^i\deb_\rho{}^{p}\deb^{(2)}_{kp}
\,+\, B(\g_b)_{\a\g} \deb^{(2)}{}^i{}_{k} \,-\, \fracm 12
B\d^i_k(\g^3\g_b)_{\a\g}(\g^3)^{\rho\tau}\deb^{(2)}_{\rho\tau}\Big)\Bar{U}
\cr
&~~~ ~-~\de_{\g k}J_\a{}^i{}_b ~+~ T_b{}_{\g k}{}^\d{}_lJ_\d{}^l{}_\a{}^i  ~~,
} \eqno({\rm A}.19)
$$
and progress is achieved in analyzing this identity by noting that it holds
$$  \eqalign{
\deb_\a{}^i\deb_\b{}^{k}\deb^{(2)}_{jk}\Bar{U}~=~
\Big(\hf C_{\a\b}\deb^{(2)}{}^{ik}\deb^{(2)}_{jk}
~+~ \hf C^{ik}\deb^{(2)}_{\a\b}\deb^{(2)}_{jk}
~-~ 2B(\g^3)_{\a\b}C^{ip}\deb^{(2)}_{pj}\Big)\bar{U}  ~~~.
}   \eqno({\rm A}.20)  $$
One other identity tells us
$$  \eqalign{
\deb^{(2)}_{\a\b}\deb^{(2)}_{ij}\bar{U} ~=~
-2B(\g^3)_{\a\b}\deb^{(2)}_{ij}\Bar{U}
~~~,
}   \eqno({\rm A}.21) $$
so that (A.20) becomes
$$
  \eqalign{  {~~~~~~~~~~}
\deb_\a{}^i\deb_\b{}^{k}\deb^{(2)}_{jk}\Bar{U}~&=~
\Big(\hf C_{\a\b}\deb^{(2)}{}^{ik}\deb^{(2)}_{jk}
~-~ 3B(\g^3)_{\a\b}C^{ip}\deb^{(2)}_{pj}\Big)\Bar{U} \cr
~&=~
\Big(\fracm 14 \, \d_j{}^i \, C_{\a\b}\deb^{(2)}{}^{k l}\deb^{(2)}_{k l}
~-~ 3B(\g^3)_{\a\b}C^{ip}\deb^{(2)}_{pj}\Big)\Bar{U}    \cr
~&=~
\Big(\fracm 34 \, \d_j{}^i \, C_{\a\b} \, \deb^{(4)}
~-~ 3B(\g^3)_{\a\b}C^{ip}\deb^{(2)}_{pj}\Big)\Bar{U}   
~~~.
}   \eqno({\rm A}.22) 
$$
where on the first term we have used a sequence of identities (see also the
final appendix).  The final line of (A.22) can now be substituted into (A.19)
to yield after a bit of algebra
$$  \eqalign{ {~~~~~~~}
2\ri\d^i_k(\g^c)_{\a\g}J_{bc}
~&=~ \ri\Big(\,  - \fracm 14  \ve_{bc}(\g^c)_{\a\g}\d^i_k\deb^{(4)} ~+~ \fracm 12 
B\ve_{bc}\d^i_k(\g^c)_{\a\g} (\g^3)^{\rho\tau}\deb^{(2)}_{\rho\tau} \, \Big)\Bar{U}
\cr
&~~~~
-\de_{\g k}J_\a{}^i{}_b
+T_b{}_{\g k}{}^\d{}_lJ_\d{}^l{}_\a{}^i
~~~.
} \eqno({\rm A}.23) $$
Finally this equation informs us that
$$  \eqalign{
J_{ab} ~=~  \ve_{ab}\Big( \,
-  \fracm 18 \deb^{(4)} ~
+~ \fracm 14 B(\g^3)^{\a\b}\deb^{(2)}_{\a\b} \, \Big) \Bar{U}
~+~{\rm h.c.}
} \eqno({\rm A}.24)$$

There remains one final Bianchi Identity of the form
$$  \eqalign{  {~~~}
0 ~&=~ \de_{\a i}J_{bc}-\de_{b}J_{\a i}{}_{ c}~+~ \de_cJ_{\a i}{}_b
~+~ T_{\a i b}{}^{D}J_{Dc}
~+~ T_{\a i c}{}^{D}J_{Db} 
~-~ T_{bc}{}^{\d l}J_{\d l\a i} ~~~, \cr
0~&=~ \ve^{ab}\Big(\de_{\a i}J_{ab}~+~ 2\de_{a}J_{b \a i}
~-~ 2T_{\a i a}{}^{\d l}J_{\d l}{}_b
~-~ 2T_{\a i a}{}^{\d}{}_lJ_{\d}{}^l{}_b
~-~ T_{ab}{}^{\d l}J_{\d l\a i}\Big)  ~~~.
} \eqno({\rm A}.25)  $$
To prove this identity is satisfied requires a calculation of some 
length.  The key to its satisfaction requires one final identity 
$$  \eqalign{  {~~~~~~~~~}
[\de_{\a i},\deb^{(4)}] \bar{U} &=~
\Big(~ -~ \fracm {8\ri}{3} \,
(\g^a)_{\a}{}^{\rho}\de_{a}\deb_\rho{}^{p}\deb^{(2)}_{ip}
~-~ 8\ri  B\ve_{bc}(\g^b)_\a{}^{\b}\de^c\deb_{\b i}  \cr
&~~~~~~~
+  \fracm 83 \, \phi_\a{}^\g\deb_\g{}^p\deb^{(2)}_{ip} 
~+~ 8\S_\a{}^{l}\deb^{(2)}_{il}~
\Big)\Bar{U}
  }  \eqno({\rm A}.26) $$
that is valid for the supergravity covariant derivative acting on a anti-chiral scalar
superfield such as $\bar{U}$.

Other Bianchi identities, not explicitly mentioned here, are identically solved by complex conjugation of the results obtained in this section.

\newpage
\noindent
{\Large{\bf Appendix B:  Miscellaneous Identities }}
\setcounter{equation}{0}

For the reader convenience, here we also collect some useful formulas 
used in the derivations
provided in this paper and especially in Appendix A (we remind that $\bar{U}$ is anti-chiral)
$$ \eqalign{  {~~~~~~}
\deb_\a{}^i\deb_\b{}^j\, & = ~\fracm 12C_{\a\b}\deb^{(2)}{}^{ij}~+~ \fracm 12C^{ij}\deb^{(2)}_{\a\b}
~+~ BC_{\a\b}C^{ij}\cm ~-~ B(\g^3)_{\a\b}\cy^{ij}  ~~~, 
} \eqno({\rm B}.1)   $$ 
%%%%%%%%%%%%%%%%%%%%%%%%%%%%%%%%%%%%%%
$$ \eqalign{
{[}\de_{\a i}, \deb^{(2)}_{jk}{]}\Bar{U}\, & = ~
-2\ri C_{i(j}(\g^c)_{\a}{}^{\d}\de_c\deb_{\d k)}
 \Bar{U}  ~~~, {~~~~~~~~~~~~~~~~~~~~~~~~~~~~~~~~~~~~}
}   \eqno({\rm B}.2)  $$ 
%%%%%%%%%%%%%%%%%%%%%%%%%%%%%%%%%%%%%%
$$ \eqalign{
{[}\de_{\a i}, \deb^{(2)}_{\d\rho}{]}\Bar{U}\, & = ~
\Big(
-2\ri(\g^c)_{\a(\d}\de_c\deb_{\rho)i}
-G(\g^3)_{\a(\d}(\g^3)_{\rho)}{}^\g\deb_{\g i}
+GC_{\a(\d}\deb_{\rho) i}
\cr  
%%%%%%%%%%%%%%%%%%%%%
&{~~~~~~\,} - \fracm \ri2H C_{\a(\d}(\g^3)_{\rho)}{}^\tau\deb_{\tau i}
~+~  \fracm \ri2   H(\g^3)_{\a(\d}\deb_{\rho) i}
-\ri H(\g^3)_{\d\rho}\deb_{\a i}
\Big)
\Bar{U}
~,} \eqno({\rm B}.3) $$ 
%%%%%%%%%%%%%%%%%%%%%%%%%%%%%%%%%%%%%%
$$ \eqalign{
{[}\de_{\a i},(\g^3)^{\d\rho} \deb^{(2)}_{\d\rho}{]}\Bar{U}\, & = ~
-4\ri \ve^{bc}(\g_b)_{\a}{}^{\b}\de_c\deb_{\b i}
\Bar{U} ~~~, {~~~~~~~~~~~~~~~~~~~~~~~~~~~~~~~~~~~~}
}  \eqno({\rm B}.4) $$ 
%%%%%%%%%%%%%%%%%
$$ \eqalign{
\deb_{\a i}\deb^{(2)}_{jk}\Bar{U}\, & = ~
- \fracm 13  C_{i(j}\deb_\a{}^{p}\deb^{(2)}_{k)p}\Bar{U}
~~~, {~~~~~~~~~~~~~~~~~~~~~~~~~~~~~~~~~~~~~~~~~}
}  \eqno({\rm B}.5) $$ 
%%%%%%%%%%%%%%%%%%%%%%%%%%%%%%%%%%%%%%
$$ \eqalign{
\deb_{\a i}\deb^{(2)}_{\b\g}\Bar{U}\, & = ~
{1\over 3}C_{\a(\b}\deb_{\g)}{}^p\deb^{(2)}_{ip}\Bar{U}
-  \fracm 43 BC_{\a(\b}(\g^3)_{\g)}{}^\d \deb_{\d i}\Bar{U}
+  \fracm 13  B(\g^3)_{(\a\b}\deb_{\g)i}\Bar{U}~, 
}  \eqno({\rm B}.6) $$ 
%%%%%%%%%%%%%%%%%%%%%%%%%%%%%%%%%%%%%%
$$ \eqalign{
(\g^3)^{\b\g}\deb_{\a i}\deb^{(2)}_{\b\g}\Bar{U}\, & = ~
- \fracm 23  (\g^3)_\a{}^\g\deb_{\g}{}^p\deb^{(2)}_{ip}\Bar{U}
~~~, {~~~~~~~~~~~~~~~~~~~~~~~~~~~~~~~~~~~~~~~~}
}  \eqno({\rm B}.7) $$ 
%%%%%%%%%%%%%%%%%%%%%%%%%%%%%%%%%%%%%%
$$ \eqalign{
\deb^\g{}_i\deb^{(2)}_{\a\g}\Bar{U}\, & = ~
-\deb_{\a}{}^p\deb^{(2)}_{ip}\Bar{U}
+4B(\g^3)_{\a}{}^\d \deb_{\d i}\Bar{U}
~~~, {~~~~~~~~~~~~~~~~~~~~~~~~~~~~~~~~~~~~}
}  \eqno({\rm B}.8)   $$ 
%%%%%%%%%%%%%%%%%%%%%%
$$ \eqalign{
&\{\de_{\a i},\deb_\d{}^{p}\deb^{(2)}_{kp} \}\Bar{U}=
\Bigg(
3\ri(\g^a)_{\a\d}\de_a\deb^{(2)}_{ik}
-3\ri C_{ik}(\g^a)_{\a}{}^{\rho}\de_a\deb^{(2)}_{\d\rho} \cr  
%%%%%%%%%%%%%%%%%%%%%
&{~~~~~~~~~~~~~~~~~~~~~~~~~~\,} 
+ \fracm 32  C_{ik}{\phi}^\tau{}_\d(\g^3)_{\a\tau}(\g^3)^{\rho\b}\deb^{(2)}_{\b\rho}
-3\phi_{\a\d}\deb^{(2)}_{ik}
\cr  
%%%%%%%%%%%%%%%%%%%%%
&{~~~~~~~~~~~~~~~~~~~~~~~~~~\,} 
+6C_{ik}C_{\a\d}\S^{\b p}\deb_{\b p}
+6C_{ik}(\g^3)_{\a\d}(\g^3)^{\b\g}\S_{\b}{}^p\deb_{\g p}
\Bigg)\Bar{U}
~,
}  \eqno({\rm B}.9) $$ 
%%%%%%%%%%%%%%%%%%%%%%%%%%%%%%%
$$ \eqalign{ {~}
\deb^{(2)}_{\a\b}\deb^{(2)}_{ij}\Bar{U}\, & = ~
-2B(\g^3)_{\a\b}\deb^{(2)}_{ij}\Bar{U} ~~~, {~~~~~~~~~~~~~~~\,} 
{~~~~~~~~~~~~~~~\,} 
}  \eqno({\rm B}.10) $$ 
%%%%%%%%%%%%%%%%%%%%%%%%%%%%%%%%%%%
$$ \eqalign{
\deb^{(2)}{}^{\a\b}\deb^{(2)}_{\a\b}\Bar{U}\, & = ~
-\deb^{(2)}{}^{ij}\deb^{(2)}_{ij}\Bar{U}
-4B(\g^3)^{\a\b}\deb^{(2)}_{\a\b}\Bar{U} ~~~, {~~~~~~~~~~\,} 
{~~~~~~~~~~~~~~~~~~\,} 
}  \eqno({\rm B}.11) $$ 
%%%%%%%%%%%%%%%%%%%%%%%%%%%%%%%%%%%
$$ \eqalign{
\deb^{(4)}\Bar{U}
\, & := ~
-\fracm 13 \deb^{(2)}{}^{kl}\deb^{(2)}_{kl}\Bar{U}
~~~,  {~~~~~~~~~~~~~~~~~~~~~~\,} 
{~~~~~~~~~~~~~\,} 
}  \eqno({\rm B}.12) $$ 
%%%%%%%%%%%%%%%%%%%%%%%%%%%%%%%%%%%
$$ \eqalign{
\deb_\a{}^i\deb_\b{}^{k}\deb^{(2)}_{jk}\Bar{U}\, & = ~
\Big( \, \fracm 34 C_{\a\b}\d^i_j\deb^{(4)}
-3B(\g^3)_{\a\b}\deb^{(2)}{}^i{}_{j}
\Big)\Bar{U}
~~~,  {~~~~~~~~~~~~~~~~~\,} 
{~~~~~~~~~~\,} 
} \eqno({\rm B}.13) $$ 
%%%%%%%%%%%%%%%%%%%%%%%%%%%%%%%%%%%
$$ \eqalign{
\deb_{\a}{}^i\big(\deb^{(2)}{}^{\g\d}-2{B}(\g^3)^{\g\d}\big)\deb^{(2)}_{\g\d}\Bar{U}\, ~= ~0~,
} 
{~~~~~~~~~~~~~~~\,} 
{~~~~~~~~~~~~~~~\,} 
 \eqno({\rm B}.14) $$ 
%%%%%%%%%%%%%%%%%%%%%%%%%%%%%%%%%
$$ \eqalign{
[\de_{\a i},\deb^{(4)}]\Bar{U}\, & = ~
\Bigg(
- \fracm {8\ri}{3} (\g^c)_{\a}{}^{\b}\de_c\deb_\b{}^k\deb^{(2)}_{ik}
-8\ri B\ve^{ab}(\g_a)_\a{}^\d\de_b \deb_{\d i}  \cr
%%%%%%%%%%%%%
&{~~~~~~~~} +8\S_\a{}^{j}\deb^{(2)}_{ij} + \fracm 83  \phi_{\a}{}^{\g}\deb_\g{}^k\deb^{(2)}_{ik}
 \Bigg)\Bar{U}
 ~.~~~~~~~~~
}  \eqno({\rm B}.15) $$ 
By complex conjugation, the reader can derive an analogue set of equations for the chiral superfield $U$.

\end{document}